\documentstyle[12pt]{article}
\newcommand{\doublespace}{\renewcommand{\baselinestretch}{1.75}
\Large\normalsize}

\begin{document}
\doublespace
\begin{titlepage}

\centerline{\bf HAMILTONIAN FORMULATION OF GENERAL RELATIVITY} 
\centerline{\bf IN THE TELEPARALLEL GEOMETRY}
\vskip 1.0cm
\bigskip
\centerline{\it J. W. Maluf$\,^{*}$ and J. F. da Rocha-Neto}
\centerline{\it Instituto de F\'isica}
\centerline{\it Universidade de Bras\'ilia}
\centerline{\it C.P. 04385}
\centerline{\it 70.919-970  Bras\'ilia, DF}  
\centerline{\it Brazil}
\date{}
\begin{abstract}
We establish the Hamiltonian formulation of the teleparallel
equivalent of general relativity, without fixing the time gauge
condition, by rigorously performing the Legendre transform. The
time gauge condition, previously considered, restricts the
teleparallel geometry to the three-dimensional spacelike
hypersurface. Geometrically, the teleparallel geometry is now
extended to the four-dimensional space-time. The resulting
Hamiltonian formulation is structurally different from the
standard ADM formulation in many aspects, the main one being
that the dynamics is now governed  by the Hamiltonian constraint
$H_0$ and a set of primary constraints. The vector constraint
$H_i$ is {\it derived} from the Hamiltonian constraint. The
vanishing of the latter implies the vanishing of the vector
constraint.

\end{abstract}
\thispagestyle{empty}
\vfill
\par
\noindent (*) e-mail: wadih@fis.unb.br
\end{titlepage}
\newpage

\noindent {\bf I. Introduction}\par
\bigskip
\noindent Hamiltonian formulations, when consistently
established, not only guarantee that field quantities have a well
defined time evolution, but also allow us to understand physical
theories from a different perspective.
We have learned from the work of Arnowitt, Deser and
Misner (ADM)\cite{ADM} that the Hamiltonian analysis of Einstein's
general relativity reveals the intrinsic structure of the theory:
the time evolution of field quantities  is determined by the
Hamiltonian and vector constraints. Thus four of the ten
Einstein's equations acquire a prominent status in the
Hamiltonian framework. Ultimately this
is an essential feature for the canonical approach to the
quantum theory of gravity. 

It is the case in general relativity that two distinct
Lagrangian formulations that yield Einstein's  equations lead
to completely different Hamiltonian
constructions. An important example in this respect is the
reformulation of the ordinary variational principle, based
on the Hilbert-Einstein action, in terms of self-dual
connections that define Ashtekar variables\cite{Ashtekar}.
Under a Palatini type variation of the action
integral constructed out of these field quantities
one obtains precisely Einstein's equations. Interesting
features of this approach reside in the Hamiltonian domain.

Einstein's general relativity can also be reformulated in the
context of the teleparallel (Weitzenb\"ock)
geometry\cite{Weitzenbock}. In this geometrical
setting the dynamical field quantities correspond to
orthornormal tetrad fields $e^a\,_\mu$ ($ a,\,\mu $ are
SO(3,1) and space-time indices, respectively). These fields 
allow the construction of the Lagrangian density of the
teleparallel equivalent of general relativity (TEGR)
\cite{Mol,Hehl,Hay,Kop,Muller,BM,Nes,Per,Maluf1},
which offers an alternative geometrical
framework for Einstein's equations. The Lagrangian density
for the tetrad field in the TEGR is given by a sum of quadratic
terms in the torsion tensor $T^a\,_{\mu \nu} =
\partial_\mu e^a\,_\nu - \partial_\nu e^a\,_\mu$, which is
related to the anti-symmetric part of Cartan's connection
$\Gamma^\lambda _{\mu \nu}= e^{a\lambda}\partial_\mu e_{a\nu}$.
The curvature tensor constructed out of the latter
vanishes identically. This connection defines a space with
teleparallelism, or absolute
parallelism\cite{Schouten}.

In a space-time with an underlying tetrad field two  vectors at
distant points are called parallel\cite{Mol} if they have
identical components with respect to the local tetrads at
the points considered. Thus consider a vector field $V^\mu(x)$.
At the point $x^\lambda$ its tetrad components are given by
$V^a(x)=e^a\,_\mu(x)V^\mu(x)$. For the tetrad components
$V^a(x+dx)$ it is easy to show that $V^a(x+dx)=V^a(x)+DV^a(x)$,
where $DV^a(x)=e^a\,_\mu(\nabla_\lambda V^\mu)dx^\lambda$. 
The covariant derivative $\nabla$ is constructed out of 
Cartan's connection
$\Gamma^\lambda_{\mu \nu}=e^{a\lambda}\partial_\mu e_{a\nu}$.
Therefore the vanishing of such covariant derivative defines
a condition for absolute parallelism in space-time.
Hence in the teleparallel geometry tetrad fields transform
under the global SO(3,1) group.
Teleparallel geometry is less restrictive than Riemannian
geometry\cite{Nester}. For a given Riemaniann geometry there
are many ways to construct  the teleparallel geometry,  since
one Riemaniann geometry corresponds to a whole equivalence
class of teleparallel geometries.

In the framework of the TEGR it is possible to make definite
statements about the energy and momentum of the gravitational
field. This fact constitutes the major motivation for
considering this theory. In the 3+1 formulation of the
TEGR\cite{Maluf1}, and by imposing Schwinger's time gauge
condition\cite{Schwinger}, we find that the
Hamiltonian and vector constraints contain each one a divergence
in the form of scalar and vector densities, respectively,
that can be identified with the energy and momentum
{\it densities} of the gravitational field\cite{Maluf2}.

In this paper we carry out the Hamiltonian formulation
of the TEGR  without
imposing the time gauge condition, by rigorously performing
the Legendre transform. We have not found it necessary
to establish a 3+1 decomposition for the tetrad field.
We only assume $g^{00}\ne 0$, a condition that ensures
that $t=constant$ hypersurfaces are spacelike. The
Lagrange multipliers are given by the zero components of the
tetrads, $e_{a0}$. The constraints corresponding to the 
Hamiltonian ($H_0$) and vector ($H_i$) constraints are obtained
in the form $C^a=0$. The dynamical evolution of the
field quantities is completely determined by $H_0$ and by a set
of primary constraints $\Gamma^{ik}$ and $\Gamma^k$, as we
will show. The surprising feature is that if $H_0=0$
in the subspace of the phase space determined by
$\Gamma^{ik}=\Gamma^k=0$,  then it follows
that $H_i=0$. As we will see, $H_i$ can be obtained from
the very definition of $H_0$. Furthermore by calculating
Poisson brackets we show that the constraints constitute a
first class set. Hence the theory is well defined regarding
time evolution.

As a consequence of this analysis, we arrive at a scalar
density that transforms as a four-vector in the SO(3,1)
space, again arising in the expression of the constraints
of the theory, and whose zero component is related
to the energy of the gravitational field. In analogy with
previous investigations, we interpret the constraint
equations $C^a=0$ as energy-momentum equations for the
gravitational field. 

The analysis developed here is similar to that
developed in Ref. \cite{Maluf5}, in which the Hamiltonian
formulation of the TEGR in null surfaces was established.
The 3+1 formulation of the TEGR has already been considered
in Ref. \cite{Nes}. There are several differences between
the latter and the present analysis. The investigation in
Ref. \cite{Nes} has not pointed out neither
the emergence of the scalar densities mentioned above
nor the relationship between $H_0$ and $H_i$. Our
approach is different and allowed us to proceed further
in the understanding of the constraint structure of the
theory.\par

\bigskip
\noindent Notation: spacetime indices $\mu, \nu, ...$ and SO(3,1)
indices $a, b, ...$ run from 0 to 3. Time and space indices are
indicated according to
$\mu=0,i,\;\;a=(0),(i)$. The tetrad field $e^a\,_\mu$ 
yields the definition of the torsion tensor:  
$T^a\,_{\mu \nu}=\partial_\mu e^a\,_\nu-\partial_\nu e^a\,_\mu$.
The flat, Minkowski spacetime  metric is fixed by
$\eta_{ab}=e_{a\mu} e_{b\nu}g^{\mu\nu}= (-+++)$.        \\

\bigskip
\bigskip

\noindent {\bf II. Lagrangian formulation}\par
\bigskip

In order to carry out the 3+1 decomposition we need a first
order differential formulation of the Lagrangian density of
the TEGR. For this purpose we
introduce an auxiliary field quantity $\phi_{abc}=-\phi_{acb}$
that will be related to the torsion tensor.
The first order differential Lagrangian formulation 
in empty space-time reads

$$L(e,\phi)\;=\;
k\,e\,\Lambda^{abc}(\phi_{abc}\,-\,2T_{abc})\;,\eqno(1)$$

\noindent where $T_{abc}=e_b\,^\mu e_c\,^\nu T_{a\mu\nu}$. 
$\Lambda^{abc}$ is defined by

$$\Lambda^{abc}\;=\;
{1\over 4}(\phi^{abc}+\phi^{bac}-\phi^{cab})+
{1\over 2}(\eta^{ac}\phi^b-\eta^{ab}\phi^c)\;,\eqno(2)$$

\noindent and $\phi_b=\phi^a\,_{ab}$.
The Lagrangian density (1) is invariant 
under coordinate and global SO(3,1) transformations.

Variation of the action constructed out of (1)
with respect to $\phi^{abc}$ yields an equation that can be
reduced to $\phi_{abc}\;=\;T_{abc}$. This equation can be
split into two equations:

$$\phi_{a0k}\;
=\;T_{a0k}\;=\;
\partial_0 e_{ak}-\partial_k e_{a0}\;,\eqno(3a)$$

$$\phi_{aik}\;
=\;T_{aik}\;=\;
\partial_i e_{ak}-\partial_k e_{ai}\;.\eqno(3b)$$

\noindent The variation of the action
integral with respect to $e_{a\mu}$ yields the field equation

$${{\delta L}\over {\delta e^{a\mu}}}\;=\;
e_{a\lambda}e_{b\mu}\partial_\nu(e\Sigma^{b\lambda \nu})-
e\biggl(\Sigma^{b \nu}\,_aT_{b\nu \mu}-
{1\over 4}e_{a\mu}T_{bcd}\Sigma^{bcd}\biggr)
\;=\;0\;.\eqno(4)$$

\noindent The tensor $\Sigma^{abc}$ is defined in terms of
$T^{abc}$ exactly like $\Lambda^{abc}$ in terms of $\phi^{abc}$.
By explicit calculations\cite{Maluf1} it is verified
that these equations are equivalent to
Einstein's equations in tetrad form:

$${{\delta L}\over {\delta e^{a\mu}}}\; \equiv \;{1\over 2}\,e\,
\biggl\{ R_{a\mu}(e)-{1\over 2}e_{a\mu}R(e)\biggr\}\;.$$

We note finally that by substituting (3a,b) into (1) the
Lagrangian density reduces to

$$L(e_{a\mu})=-k\,e \Sigma^{abc}T_{abc}=
-k\,e\,({1\over 4}T^{abc}T_{abc}+{1\over 2} T^{abc}T_{bac}
-T^aT_a)\;.$$

\bigskip
\bigskip

\noindent {\bf III. Legendre transform and the 3+1 decomposition}\par

\bigskip
The Hamiltonian density will be obtained by the standard
prescription $L=p\dot q -H_0$ and by properly identifying primary
constraints. We have not found it necessary to establish any
kind of 3+1 decomposition for the tetrad fields. Therefore in
the following both $e_{a\mu}$ and $g_{\mu \nu}$ are space-time
fields. We will follow here the procedure
presented in \cite{Maluf5}.

Lagrangian density (1) can be expressed as

$$L(e,\phi)\;=\;-4ke\,\Lambda^{a0k}\,\dot e_{ak}+
4ke\,\Lambda^{a0k}\,\partial_k e_{a0}
-2ke\,\Lambda^{aij}\,T_{aij}
+ke\Lambda^{abc}\,\phi_{abc}\;,\eqno(5)$$

\noindent where the dot indicates time derivative, and
$\Lambda^{a0k}=\Lambda^{abc}\,e_b\,^0\,e_c\,^k$,
$\Lambda^{aij}=\Lambda^{abc}\,e_b\,^i\,e_c\,^j$.

\noindent Therefore the momentum canonically conjugated to
$e_{ak}$ is given by

$$\Pi^{ak}\;=\;-4k\,e\,\Lambda^{a0k}\;,\eqno(6)$$

\noindent In terms of (6) expression (5) reads

$$L\;=\; \Pi^{ak}\,\dot e_{ak}-\Pi^{ak}\,\partial_k e_{a0}-
2ke\,\Lambda^{aij}\,T_{aij}+
ke\,\Lambda^{abc}\,\phi_{abc}$$

$$=  \Pi^{ak}\,\dot e_{ak}-\Pi^{ak}\,\partial_k e_{a0}-
ke\Lambda^{aij}(2T_{aij}-   \phi_{aij})
+2ke\Lambda^{a0k}\phi_{a0k}\;.\eqno(7)$$

\noindent The last term on the right hand side of equation (7)
is identified as $2ke\Lambda^{a0k}\phi_{a0k}=
-{1\over 2}\Pi^{ak}\phi_{a0k}$. 

The Hamiltonian formulation is established once we rewrite the
Lagrangian density (7) in terms of $e_{ak}$, $\Pi^{ak}$ and
further nondynamical field quantities. It is carried out in
two steps. First, we take into account equation (3b) in (7)
so that half of the auxiliary fields, $\phi_{aij}$, are
eliminated from the Lagrangian by means of the identification

$$\phi_{aij}=T_{aij}\;.$$

\noindent As a consequence we have

$$- ke\Lambda^{aij}(2T_{aij}-   \phi_{aij})=
- ke\Lambda^{aij} T_{aij}$$

$$=-ke\biggl( {1\over 4}g^{im}g^{nj} T^a\,_{mn}T_{aij}+
{1\over 2}g^{nj} T^i\,_{mn}T^m\,_{ij}-
g^{ik}T^j\,_{ji}T^n\,_{nk}\biggr)$$

$$+ke\biggl(-{1\over 2}g^{0i}g^{jk}\phi^a\,_{0k}T_{aij}-
{1\over 2}g^{jk}\phi^i\,_{0k}T^0\,_{ij}+
{1\over 2}g^{0j}\phi^i\,_{0k}T^k\,_{ij}-
g^{0k}\phi^j\,_{0j}T^i\,_{ik}+
g^{ik}\phi^0\,_{0i}T^j\,_{jk}\biggr)\;.$$

\noindent The last five terms of the expression above may be
rewritten as

$$- {1\over 2}ke\,\phi_{a0k} \biggl[
g^{0i}g^{kj}T^a\,_{ij}-
e^{ai}(g^{0j}T^k\,_{ij}-g^{kj}T^0\,_{ij})
+2(e^{ak}g^{0i}-e^{a0}g^{ki})T^j\,_{ji}\biggr]
\;.$$

\noindent Therefore we have

$$L(e_{ak},\Pi^{ak},e_{a0},\phi_{a0k})\;
=\;\Pi^{ak}\dot e_{ak}+e_{a0}\partial_k\Pi^{ak}-
\partial_k(e_{a0}\Pi^{ak})$$

$$-ke\biggl( {1\over 4}g^{im}g^{nj}T^a\,_{mn}T_{aij}+
{1\over 2}g^{nj}T^i\,_{mn}T^m\,_{ij}-
g^{ik}T^j\,_{ji}T^n\,_{nk}\biggr)$$

$$-{1\over 2}\phi_{a0k}\biggl\{ \Pi^{ak}+
ke\biggl[ g^{0i}g^{kj}T^a\,_{ij}-
e^{ai}(g^{0j}T^k\,_{ij}-g^{kj}T^0\,_{ij})
+2(e^{ak}g^{0i}-e^{a0}g^{ki})T^j\,_{ji}\biggr]
\biggr\}\;.\eqno(8)$$

The second step consists of expressing the remaining
auxiliary field quantities, the ``velocities" $\phi_{a0k}$,
in terms of the momenta $\Pi^{ak}$. This is the nontrivial
step of the Legendre transform. 

We need to consider the full
expression of $\Pi^{ ak}$. It is given by equation (6),

$$\Pi^{ak}\;=\;k\,e\biggl\{ 
g^{00}(-g^{kj}\phi^a\,_{0j}-
e^{aj}\phi^k\,_{0j}+2e^{ak}\phi^j\,_{0j})$$

$$+g^{0k}(g^{0j}\phi^a\,_{0j}+e^{aj}\phi^0\,_{0j})
\,+e^{a0}(g^{0j}\phi^k\,_{0j}+g^{kj}\phi^0\,_{0j})
-2(e^{a0}g^{0k}\phi^j\,_{0j}+e^{ak}g^{0j}\phi^0\,_{0j})$$

$$-g^{0i}g^{kj}T^a\,_{ij}+e^{ai}(g^{0j}T^k\,_{ij}-
g^{kj}T^0\,_{ij})-2(g^{0i}e^{ak}-g^{ik}e^{a0})
T^j\,_{ji} \biggr\}\;,\eqno(9)$$

\noindent where we have already identified
$\phi_{aij}=T_{aij}$.
Denoting $(..)$ and $[..]$ as the symmetric and
anti-symmetric parts  of field quantities, respectively, we
decompose $\Pi^{ak}$ into irreducible components:

$$\Pi^{ak}\;=\;e^a\,_i\,\Pi^{(ik)}+e^a\,_i\,\Pi^{[ik]}+
e^a\,_0\,\Pi^{0k}\;,\eqno(10)$$

\noindent where

$$\Pi^{(ik)}\;=\;k\,e\biggl\{
g^{00}(-g^{kj}\phi^i\,_{0j}-
g^{ij}\phi^k\,_{0j}+2g^{ik}\phi^j\,_{0j})
+g^{0k}( g^{0j}\phi^i\,_{0j}+
g^{ij}\phi^0\,_{0j}-g^{0i}\phi^j\,_{0j})$$

$$+g^{0i}(g^{0j}\phi^k\,_{0j}+
g^{kj}\phi^0\,_{0j}-g^{0k}\phi^j\,_{0j})
-2g^{ik}\,g^{0j}\phi^0\,_{0j}\;+\;\Delta^{ik}\biggr\}
\;,\eqno(11a)$$

$$\Delta^{ik}\;=\;-g^{0m}(
g^{kj}T^i\,_{mj}+g^{ij}T^k\,_{mj}-2g^{ik}T^j\,_{mj})
-(g^{km}g^{0i}+g^{im}g^{0k}) T^j\,_{mj}\;,\eqno(11b)$$

$$\Pi^{[ik]}\;=\;k\,e\biggl\{ -g^{im}g^{kj}T^0\,_{mj}+
(g^{im}g^{0k}-g^{km}g^{0i})T^j\,_{mj}\biggr\}
\;,\eqno(12)$$

$$\Pi^{0k}\; =\;-2k\,e\, (
g^{kj}g^{0i}T^0\,_{ij}-g^{0k}g^{0i}T^j\,_{ij}
+g^{00}g^{ik}T^j\,_{ij} )
\;.\eqno(13)$$

The crucial point in this analysis is that only the symmetrical
components $\Pi^{(ij)}$ depend on the ``velocities" $\phi_{a0k}$.
The other six components, $\Pi^{\lbrack ij \rbrack}$ and
$\Pi^{0k}$ depend solely on $T_{aij}$. Therefore we can
express only six of the ``velocity" fields $\phi_{a0k}$  in
terms of the components
$\Pi^{(ij)}$. With the purpose of finding out which
components of $\phi_{a0k}$ can be inverted in terms of the
momenta we decompose $\phi_{a0k}$ identically as

$$\phi^a\,_{0j}\;=\;e^{ai}\,\psi_{ij}+e^{ai}\,\sigma_{ij}
+e^{a0}\,\lambda_j\;,\eqno(14)$$

\noindent where
$\psi_{ij}=
{1\over 2}(\phi_{i0j}+\phi_{j0i})$,$\;\;$ 
$\sigma_{ij}=
{1\over 2}(\phi_{i0j}-\phi_{j0i})$,$\;\;$
$\lambda_j=\phi_{00j}$,$\;\;$
and $\phi_{\mu 0j}=e^a\,_\mu \phi_{a0j}$
(like $\phi_{abc}$, the components $\psi_{ij}$,
$\sigma_{ij}$ and $\lambda_j$ are also auxiliary field
quantities). Next we substitute (14) in (11a). By defining

$$P^{ik}\;=\;{1\over {ke}}\Pi^{(ik)}-\Delta^{ik}\;,\eqno(15)$$

\noindent we find that $P^{ik}$ depends only on $\psi_{ij}$:

$$P^{ik}\;=\;-2g^{00}(g^{im}g^{kj}\psi_{mj}-g^{ik}\psi)$$

$$+2(g^{0i}g^{km}g^{0j}+g^{0k}g^{im}g^{0j})\psi_{mj}
-2(g^{ik}g^{0m}g^{0j}\psi_{mj}+g^{0i}g^{0k}\psi)\;,\eqno(16)$$

\noindent where $\psi=g^{mn}\psi_{mn}$.

We can now invert $\psi_{mj}$ in terms of $P^{ik}$. After a
number of manipulations we arrive at

$$\psi_{mj}\;=\;-{1\over{2g^{00}}}\biggl(
g_{im}g_{kj}P^{ik}-{1\over 2}g_{mj}\,P\biggr)\;,\eqno(17)$$

\noindent where $P=g_{ik}P^{ik}$.

At last we  need to rewrite the third line of the Lagrangian
density (8) in terms of canonical variables.
By making use of (9), (14) and (17) we can rewrite

$$-{1\over 2}\phi_{a0k}\biggl\{ \Pi^{ak}+
ke\biggl[ g^{0i}g^{kj}T^a\,_{ij}-
e^{ai}(g^{0j}T^k\,_{ij}-g^{kj}T^0\,_{ij})
+2(e^{ak}g^{0i}-e^{a0}g^{ki})T^j\,_{ji}\biggr]
\biggr\}$$

\noindent in the form

$${1\over {4g^{00}}}ke\biggl(
g_{ik}g_{jl}P^{ij}P^{kl}-{1\over 2}P^2\biggr)\;.$$

Thus we finally obtain the primary Hamiltonian density
$H_0=\Pi^{ak}\dot e_{ak} -L$,

$$H_0(e_{ak}, \Pi^{ak}, e_{a0})\;=\;-e_{a0}\partial_k \Pi^{ak}
-{1\over {4g^{00}}} ke \biggl(g_{ik}g_{jl}P^{ij}P^{kl}-
{1\over 2}P^2\biggr)$$

$$+ke\biggl( {1\over 4}g^{im}g^{nj}T^a\,_{mn}T_{aij}
+{1\over 2}g^{nj}T^i\,_{mn}T^m\,_{ij}
-g^{ik}T^j\,_{ji}T^n\,_{nk}\biggr)\;.\eqno(18)$$

We may now write the total Hamiltonian density. For this purpose
we have to identify the primary constraints. They are given by
expressions (12) and (13), which represent relations between
$e_{ak}$ and the momenta $\Pi^{ak}$. Thus we define

$$\Gamma^{ik}\;=\;-\Gamma^{ki}\;=\;
\Pi^{[ik]}\;-\;k\,e\biggl\{ -g^{im}g^{kj}T^0\,_{mj}+
(g^{im}g^{0k}-g^{km}g^{0i})T^j\,_{mj}\biggr\}
\;,\eqno(19)$$

$$ \Gamma^k \;=\;\Pi^{0k}\; +\;2k\,e\, (
g^{kj}g^{0i}T^0\,_{ij}-g^{0k}g^{0i}T^j\,_{ij}
+g^{00}g^{ik}T^j\,_{ij} )
\;.\eqno(20)$$

\noindent Therefore the total Hamiltonian density is given by

$$H(e_{ak},\Pi^{ak},e_{a0},\alpha_{ik},\beta_k)
=H_0 + \alpha_{ik}\Gamma^{ik} +\beta_k \Gamma^k
+\partial_k(e_{a0}\Pi^{ak}) \;,\eqno(21)$$

\noindent where $\alpha_{ik}$ and $\beta_k$ are Lagrange
multipliers.\par
\bigskip
\bigskip

\noindent {\bf IV. Secondary constraints}\par

\bigskip
Since the momenta $\lbrace \Pi^{a0} \rbrace$  vanish
identically they also constitute primary constraints that induce
the secondary constraints

$$C^a\equiv {{\delta H}\over {\delta e_{a0}}}=0\;.\eqno(22)$$

\noindent In order to obtain the expression of $C^a$ we have only
to vary $H_0$ with respect to $e_{a0}$, because
variations of $\Gamma^{ik}$ and
$\Gamma^k$ with respect to $e_{a0}$ yield the constraints
themselves:

$${{\delta \Gamma^{ik}  } \over{\delta e_{a0}}}=
-{1\over 2}(e^{ai}\Gamma^k-e^{ak}\Gamma^i)\;,\eqno(23a)$$

$${{\delta  \Gamma^k  }\over{\delta e_{a0}}}=
-e^{a0}\Gamma^k\;.\eqno(23b)$$

\noindent In (23a,b) we have made use of variations like
$\delta e^{b\mu} / \delta e_{a0}=-e^{a\mu}e^{b0}\;.$
In the process of obtaining $C^a$ we need the
variation of $P^{ij}$ with respect to $e_{a0}$. It reads

$${{\delta P^{ij}}\over{\delta e_{a0}}}=
-e^{a0}P^{ij}+\gamma^{aij}\;,$$

\noindent with $\gamma^{aij}$  defined by

$$\gamma^{aij}\;=\;-{1\over {2ke}}(e^{ai}\Gamma^j+
e^{aj}\Gamma^i)
-e^{ak}\biggl[
g^{00}( g^{jm}T^i\,_{km}+g^{im}T^j\,_{km}+
2g^{ij}T^m\,_{mk})$$

$$+g^{0m}(g^{0j}T^i\,_{mk}+g^{0i}T^j\,_{mk})
-2g^{0i}g^{0j}T^m\,_{mk}
+(g^{jm}g^{0i}+g^{im}g^{0j}-2g^{ij}g^{0m})T^0\,_{mk}
\biggr]\;.\eqno(24)$$

\noindent Note that $\gamma^{aij}$ satisfies
$e_{a0}\gamma^{aij}= 0$.

After a long calculation we arrive at the expression of $C^a$:

$$C^a\;=\;-\partial_k \Pi^{ak}+
e^{a0}\biggl[-{1\over{4g^{00}}}ke\biggl(g_{ik}g_{jl}P^{ij}P^{kl}-
{1\over 2}P^2\biggr)$$

$$+ke\biggl( {1\over 4}g^{im}g^{nj}T^b\,_{mn}T_{bij}+
{1\over 2}g^{nj}T^i\,_{mn}T^m\,_{ij}
-g^{ik}T^m\,_{mi}T^n\,_{nk}\biggr) \biggr]$$

$$-{1\over {2g^{00}}}ke\biggl(g_{ik}g_{jl}\gamma^{aij}P^{kl}-
{1\over 2}g_{ij}\gamma^{aij}\,P\biggr)
-ke\,e^{ai}\biggl(g^{0m}g^{nj}T^b\,_{ij}T_{bmn}$$

$$+g^{nj}T^0\,_{mn}T^m\,_{ij}+g^{0j}T^n\,_{mj}T^m\,_{ni}
-2g^{0k}T^m\,_{mk}T^n\,_{ni}-2g^{jk}T^0\,_{ij}T^n\,_{nk}
\biggr)\;.\eqno(25)$$

Inspite of the fact that expression above is somehow intricate,
we immediately notice that

$$e_{a0}C^a=H_0\;.\eqno(26)$$

\noindent Therefore the total Hamiltonian becomes

$$H(e_{ak},\Pi^{ak},e_{a0},\alpha_{ik},\beta_k)
=e_{a0}C^a+\alpha_{ik}\Gamma^{ik}+\beta_k \Gamma^k +
\partial_k(e_{a0}\Pi^{ak})\;.\eqno(27)$$

\noindent We observe that $\lbrace e_{a0}\rbrace$ arise as 
Lagrange multipliers (see equation (50) ahead).

Before closing this section we remark that the Hamiltonian
formulation described here is different from that developed
in Ref. \cite{Nes}, the difference residing in the definition
of the canonical momentum. In the latter reference the canonical
momentum is not defined by taking the variation of $L$ with
respect to $\dot e_{ak}$. Instead, it is defined by

$$\pi_a\,^k={{\delta L}\over{\delta (N^\bot T^a\,_{\bot k})}}=
{{\delta L}\over
{\delta(T^a\,_{0k}-N^iT^a\,_{ik})}}\;,$$

\noindent where $N^\bot$ and $N^i$ are the usual lapse and
shift functions.
As a consequence, three of the six primary constraints of
Ref. \cite{Nes} are different from the corresponding
constraints obtained here.
The expression of the components $\tau^{\lbrack ik \rbrack}$
and $\tau_{\bot}\,^k$ 
of Ref. \cite{Nes}, equivalent to  $\Pi^{\lbrack ik \rbrack}$
and $\Pi^{0k}$, respectively, given by (12) and (13), read
in our notation

$$\tau^{\lbrack ik \rbrack}=
-e \biggl\{ g^{im}g^{kj}T^0\,_{ij}+
N^j(g^{im}g^{0k}-g^{km}g^{0i})T^0\,_{mj} \biggr\} \;,$$

$$\tau_{\bot}\,^k={1\over {2k}}N^\bot\,\Pi^{0k}\;.$$

\noindent The Hamiltonian and vector constraints of the above
mentioned reference are parametrized in terms of the lapse
and shift functions. In
the present work we have parametrized the set of four
constraints according to equation (26), and identified the
Lagrange multipliers as $e_{a0}$. The final expression of
$C^a$ acquires  the total divergence $-\partial_k \Pi^{ak}$.
This divergence is different
from the one that appears in the
expression of the total Hamiltonian density of gravitational
fields for asymptotically flat space-times, either in the
metric\cite{Regge} or in the tetrad formulation (see, for
example, Eq. (3.17) of Ref. \cite{Nes} or Eq. (27) above;
it is possible to show that the latter expressions are
exactly the same field quantities). We finally notice
that the constraint algebra to be presented in
the coming section has not been evaluated in Ref. \cite{Nes}.

\bigskip
\bigskip
\vskip 1.0cm
\noindent {\bf V. Simplification of the constraints
and Poisson brackets}\par
\bigskip
The first two terms of the expression of $C^a$ yield
the primary Hamiltonian in the form $e^{a0}H_0$. This fact can
be easily verified by expressing the first term of (25) as

$$-\partial_k \Pi^{ak}=e^{a0}(-e_{b0}\partial_k \Pi^{bk})+
e^{aj}(-e_{bj}\partial_k \Pi^{bk})\;.$$

\noindent The second term considered above is the collection
of terms in (25) multiplied by $e^{a0}$.
Substituting definitions (11b) and (24)
for $\Delta^{ij}$ and $\gamma^{aij}$, respectively, into (25)
we obtain after a long calculation a simplified form for
$C^a$,

$$C^a=e^{a0}H_0+e^{ai}F_i\;,\eqno(28)$$

\noindent with the following definitions:

$$F_i=H_i+\Gamma^m T_{0mi}+\Gamma^{lm}T_{lmi}+
{1\over {2g^{00}}}(g_{ik}g_{jl}P^{kl}-
{1\over 2}g_{ij}P)\Gamma^j\;,\eqno(29)$$

$$H_i=-e_{bi}\partial_k \Pi^{bk}-\Pi^{bk} T_{bki}\;.\eqno(30)$$

We denote $H_0$ the Hamiltonian constraint.
$H_i$ is the vector constraint.
It amounts to a SO(3,1) version of the vector constraint of
Ref. \cite{Maluf1}. The true constraints of the theory are
$C^a$, $\Gamma^{ik}$ and $\Gamma^k$. Dispensing with the
surface term the total Hamiltonian reads

$$H=e_{a0}C^a+\alpha_{ik}\Gamma^{ik}+\beta_k\Gamma^k
\;.\eqno(31)$$

The Poisson bracket between two
quantites $F$ and $G$ is defined by

$$\lbrace F,G\rbrace=\int d^3x \biggl(
{{\delta F}\over {\delta e_{ai}(x)}}
{{\delta G}\over {\delta\Pi^{ai}(x)}}-
{{\delta F}\over {\delta\Pi^{ai}(x)}}
{{\delta G}\over {\delta e_{ai}(x)}} \biggr)\;,$$

\noindent by means of which we can write down the evolution
equations. The first set of Hamilton's equations is given by

$$\dot e_{aj}(x)= \lbrace e_{aj}(x), {\bf H}\rbrace =\int d^3y
{{\delta }\over{\delta\Pi^{aj}(x)}}
\biggl( H_0(y)+\alpha_{ik}(y)\Gamma^{ik}(y)
+\beta_k(y)\Gamma^k(y)\biggr) \;,\eqno(32)$$

\noindent where {\bf H} is the total Hamiltonian. 
This equation can be worked to yield

$$T_{a0j}=  
-{1\over {2g^{00}}}e_a\,^k(g_{ik}g_{jm}P^{im}-
{1\over 2}g_{kj}P) +e_a\,^i\alpha_{ij}+
e_a\,^0\beta_j\;,\eqno(33)$$

\noindent from which we obtain

$${1\over 2} (T_{i0j}+T_{j0i})=\psi_{ij}=
-{1\over {2g^{00}}}
(g_{ik}g_{mj}P^{km}-{1\over 2}g_{ij}P)\;,\eqno(34a)$$

$${1\over 2} (T_{i0j}-T_{j0i})=\sigma_{ij}
=\alpha_{ij}\;,\eqno (34b)$$

$$T_{00j}=\lambda_j=\beta_j\;,\eqno(34c)$$

\noindent according to the definitions in equation (14).
Thus the Lagrange multipliers in (31) acquire a
well defined meaning. Expression (34a) is in total agreement
with (17). Consequently we can obtain an expression for
$\Pi^{(ij)}$ in terms of velocities via equations (15) and
(16). The dynamical evolution of the field quantities is
completed with Hamilton's equations for $\Pi^{(ij)}$,

$$\dot \Pi^{(ij)}(x)=\lbrace \Pi^{(ij)}(x), {\bf H}\rbrace=
\int d^3y \biggl(
{{\delta \Pi^{(ij)}(x)}\over {\delta e_{ak}(y)}}
{{\delta {\bf H}}\over {\delta \Pi^{ak}(y)}}-
{{\delta \Pi^{(ij)}(x)}\over {\delta \Pi^{ak}(y)}}
{{\delta {\bf H}}\over {\delta e_{ak}(y)}}
\biggr)\;,\eqno(35)$$

\noindent together with

$$\Gamma^{ik}\;=\; \Gamma^k=0\;.\eqno(36)$$

The calculations of the Poisson brackets between
these  constraints are exceedingly
complicated. Here we will just present the results.
Instead of considering $C^a(x)$ in the calculations below, we
found it more appropriate to consider $H_0(x)$ and $H_i(x)$.
The constraint algebra is given by

$$\lbrace H_0(x),H_0(y)\rbrace=0\;,\eqno(37)$$

$$\lbrace H_0(x),H_i(y)\rbrace=-H_0(x)
{\partial\over{\partial y^i}}\delta(x-y)$$

$$-H_0e^{a0}\partial_ie_{a0}\delta(x-y)-
F_j e^{aj}\partial_ie_{a0}\delta(x-y)
\;,\eqno(38)$$

$$\lbrace H_j(x),H_k(y)\rbrace=
-H_k(x){\partial \over {\partial x^j}}\delta(x-y)-
H_j(y){\partial \over {\partial y^k}}\delta(x-y)\;,\eqno(39)$$

$$\lbrace \Gamma^i(x),\Gamma^j(y)\rbrace=0\;,\eqno(40)$$

$$\lbrace \Gamma^{ij}(x),\Gamma^k(y)\rbrace =
(g^{0j}\Gamma^{ki}-g^{0i}\Gamma^{kj})\delta(x-y)\;,\eqno(41)$$

$$\lbrace \Gamma^{ij}(x),\Gamma^{kl}(y)\rbrace={1\over 2}\biggl(
g^{il}\Gamma^{jk}+g^{jk}\Gamma^{il}-
g^{ik}\Gamma^{jl}-g^{jl}\Gamma^{ik}\biggr)\delta(x-y)
\;,\eqno(42)$$

$$\lbrace H_0(x),\Gamma^{ij}(y)\rbrace=
\biggl[ {1\over{2g^{00}}}P^{kl}\biggl( {1\over 2}
g_{kl}g_{mn}-g_{km}g_{nl}\biggr)\biggl(
g^{mi}\Gamma^{nj}-g^{mj}\Gamma^{ni}\biggr)+$$

$$+{1\over 2}\biggl(\Gamma^{nj}e^{ai}-\Gamma^{ni}e^{aj}\biggr)
\partial_n e_{a0}\biggr]\delta(x-y)\;,\eqno(43)$$

$$\lbrace H_0(x),\Gamma^i(y)\rbrace =\biggl[g^{0i}H_0+
{1\over {g^{00}}}P^{kl}\biggl({1\over 2}g_{kl}g_{jm}-
g_{kj}g_{ml}\biggr)g^{0j}\Gamma^{mi}$$

$$+\biggl( \Gamma^{ni}e^{a0} 
+\Gamma^n e^{ai}\biggr)  \partial_n e_{a0}+
{1\over 2}\Gamma^{mn}T^i\,_{nm}$$

$$+2\partial_n \Gamma^{ni}+g^{in}\biggl( H_n-\Gamma^jT_{0nj}-
\Gamma^{mj}T_{mnj}\biggr)\biggr]\delta(x-y)$$

$$+\Gamma^{ni}(x)
{\partial \over {{\partial x^n}}}\delta(x-y)\;,\eqno(44)$$

$$\lbrace H_i(x),\Gamma^j(y)\rbrace=
\delta^j_i\Gamma^n(y){\partial\over{\partial y^n}}
\delta(x-y)+ \Gamma^j(x){\partial\over{\partial x^i}}\delta(x-y)
-\Gamma^je^{a0} \partial_i e_{a0}\delta(x-y)\;,\eqno(45)$$

$$\lbrace H_k(x), \Gamma^{ij}(y)=
\Gamma^{ij}(x){\partial\over{\partial x^k}}\delta(x-y)+
\biggl( \delta^j_k \Gamma^{ni}(y)-\delta^i_k\Gamma^{nj}(y)
\biggr){\partial\over{\partial x^n}}\delta(x-y)$$

$$+{1\over 2}\biggl( e^{aj}(x)\Gamma^i(x)-
e^{ai}(x)\Gamma^j(x)\biggr) {\partial\over{\partial x^k}}
e_{a0}(x)\delta(x-y)\;.\eqno(46)$$

It is clear from the constraint algebra above that
$H_0$, $H_i$, $\Gamma^{ik}$ and $\Gamma^k$ constitute a set of
first class constraints. Now it is easy to conclude
that $C^a$, $\Gamma^{ik}$ and $\Gamma^k$  also constitute a
first class set. By means of equation (28) we have
$\lbrace C^a(x), C^b(y)\rbrace=
e^{a0}(x)\lbrace H_0(x), H_0(y) \rbrace e^{b0}(y)+$
$H_0(x) \lbrace e^{a0}(x), H_0(y) \rbrace e^{b0}(y)\,+
\cdot \cdot \cdot \cdot\;$.
On the right hand side of this Poisson bracket as well as of the
brackets $\lbrace C^a(x), \Gamma^{ik}(y)\rbrace$ and
$\lbrace C^a(x),\Gamma^k(y)\rbrace$ there will always
appear a combination of the constraints $H_0=e_{a0}C^a$,
$\Gamma^{ik}$, $\Gamma^k$ and

$$H_i=e_{ai}C^a
-\Gamma^m T_{0mi}-\Gamma^{lm}T_{lmi}-
{1\over {2g^{00}}}(g_{ik}g_{jl}P^{kl}+
{1\over 2}g_{ij}P)\Gamma^j\;.\eqno(47)$$

\noindent The expression above follows from equation (29).
All constraints of the theory are first class, and
therefore the theory is well defined regarding time
evolution.

The Hamiltonian density (31) determines the time
evolution of any field quantity $f(x)$:

$$\dot f(x)= \int d^3y \lbrace f(x), H(y)\rbrace
\biggm|_{\Gamma^{ik}=\Gamma^k=0}\;.\eqno(48)$$

\noindent Physical quantities take values in the subspace of
the phase space ${\bf P}_\Gamma$ defined by (36). In this
subspace the constraints $C^a$ become

$$C^a=e^{a0}H_0+e^{ai}H_i\;.\eqno(49)$$

Restricting considerations to ${\bf P}_\Gamma$ we
note that if $H_0$ vanishes, then
$e_{a0}C^a$ also vanishes. Since $\lbrace e_{a0}\rbrace$
are arbitrary, it follows  that $C^a=0$. In order to arrive at
this conclusion we note that the constraints
$C^a$ are independent of $e_{a0}$. From the orthogonality
relation $e_{a\mu}e^{a\lambda}=\delta_\mu^\lambda$ we obtain
$\delta e^{b\mu}/ \delta e_{a0}= -e^{a\mu}e^{b0}$.
Using this variational relation and equations (22) and (49)
it is possible to show that

$${{\delta C^a}\over{\delta e_{b0}}}={\delta \over{\delta e_{b0}}}
\biggl( e^{a0} H_0+e^{ai} H_i\biggr)=
-e^{b0}e^{a0}H_0+e^{a0}{{\delta H_0}\over{\delta e_{b0}}}-
e^{bi}e^{a0}H_i  $$

$$=-e^{b0}e^{a0}H_0+e^{a0}(e^{b0}H_0+e^{bi}H_i)-
e^{bi}e^{a0}H_i=0\;.\eqno(50)$$

\noindent $H_i$ does not depend explicitly or implicitly on
$e_{a0}$. We remark that by taking the variation with respect to
$e_{b0}$ of both sides of equation (26), $H_0=e_{a0}C^a$, we
arrive at

$$C^b=C^b+e_{a0}{{\delta C^a}\over{\delta e_{b0}}}\;,$$

\noindent from what follows the general result
$e_{a0}(\delta C^a/\delta e_{b0})=0$. Taking into account the
arbitrariness of $e_{a0}$ in the latter equation we are led to
equation (50).

Therefore the vanishing of the Hamiltonian constraint $H_0$
implies the vanishing of $C^a$, and ultimately of the vector
constraint $H_i$. Moreover we observe from (47) and (49) that
$H_i$ can be obtained from $H_0$ in ${\bf P}_\Gamma$ according to

$$e_{ai}{\delta \over {\delta e_{a0}}} H_0=
e_{ai}C^a= H_i\;.\eqno(51)$$

\noindent Thus $H_i$ is {\it derived} from $H_0$. In the complete
phase space the vanishing of $H_i$ is a consequence of the
vanishing of $H_0$, $\Gamma^{ik}$ and $\Gamma^k$.

Finally we would like to remark that
the Hamiltonian formulation of the theory can be described more
succintly in terms of the constraints $H_0$, $\Gamma^{ik}$ and
$\Gamma^k$, by the Hamiltonian density in the form

$$H(e_{ak}, \Pi^{ak}, e_{a0}, \alpha_{ik}, \beta_k)
= H_0+\alpha_{ik}\Gamma^{ik}+\beta_k\Gamma^k\;.\eqno(52)$$

\noindent The Poisson brackets between these constraints are given
by equations (37), (40-44). They constitute a first class set
except for the fact that on the right hand side of (44) there
appears the constraint $H_i$.
However it poses no problem for the consistency of the
constraints provided $H_0$, $\Gamma^{ik}$ and $\Gamma^k$ are
taken to vanish at the intial time $t=t_0$. Let $\phi(x^i,t)$
represent any of the latter constraints. At the initial time
we have $\phi(x^i,t_0)=0$. At $t_0+\delta t$ we find
$\phi(x^i,t_0+\delta t)=\phi(x^i,t_0)+\dot \phi(x^i,t_0)\delta t$
such that ${\dot \phi(x^i,t_0)}=
\lbrace \phi(x^i,t_0),{\bf H}\rbrace$. Since the vanishing of
$H_i$ at an instant of time is a consequence of the vanishing of
$H_0$, $\Gamma^{ik}$ and $\Gamma^k$ at the same time,  the
consistency of the constraints is guaranteed at any $t>t_0$.\par
\bigskip
\bigskip
\vskip 2.0cm

\noindent {\bf VI. Discussion}\par
\bigskip

The Weitzenb\"ock space-time allows a consistent description of
the Hamiltonian formulation of the gravitational field. Although
the underlying geometry is not Riemannian, the 
Lagrangian field equations (4) assure that the theory determined
by (1) is equivalent to Einstein's general relativity.
To our knowledge there does not
exist any impediment based on experimental facts that rules out
the teleparallel geometry in favour of the Riemannian geometry
for the description of the physical space-time. 
The natural geometrical setting for teleparallel gravity is the
teleparallel geometry. The Hamiltonian formulation of the TEGR
in the Riemannian geometry, with local SO(3,1) symmetry,
requires the introduction of a large number of field variables
that renders an intricate constraint structure\cite{BN}.

We have shown that the vector constraint $H_i$ can be
obtained from the Hamiltonian constraint $H_0$ by means
of a functional derivative of $H_0$, making use of the
orthogonality properties of the tetrads in the reduced
phase space ${\bf P}_\Gamma$.
However, it is an independent constraint. 
In contrast, in the ADM formulation the Hamiltonian and
vector constraints are not mutually related, and in practice
one has to consider both constraints for the dynamical
evolution via Hamilton equations.

The number of degrees of freedom may be counted as the total
number of canonical variables, $e_{ak}$ and $\Pi^{ak}$, minus
twice the number of first class constraints. Therefore we have
$24-20=4$ degrees of freedom in the phase space, as expected.
Since the constraints $\Gamma^{ik}$ and $\Gamma^k$ are first
class they act on $e_{ak}$, and $\Pi^{ak}$ and generate
symmetry transformations. In particular, for $e_{a\mu}$ we
have

$$\delta e_{ak}(x)=
\int d^3z \biggl[ \varepsilon_{ij}(z)
\lbrace e_{ak}(x), \Gamma^{ij}(z)\rbrace+
 \varepsilon_j(z)
\lbrace e_{ak}(x), \Gamma^j(z)\rbrace\biggr]$$

$$=\int d^3z \biggl[ \varepsilon_{ij}(z)
{{\delta \Gamma^{ij}(z)} \over{\delta \Pi^{ak}(x)}}+
\varepsilon_j(z)
{{\delta \Gamma^j(z)} \over{\delta \Pi^{ak}(x)}}\biggr]
= \varepsilon_{ik}e_a\,^i+
\varepsilon_k e_a\,^0\;,\eqno(53)$$

\noindent where $\varepsilon_{ij}(x)=-\varepsilon_{ji}(x)$ and
$\varepsilon_j(x)$ are infinitesimal parameters.
Note that these transformations do {\it not} act
on $e_{a0}$. This issue has not been completely analyzed.
The physical implications of these symmetries to
the theory are currently under investigation.

In the analysis of a
theory described by a Lagrangian density similar to (1),
M\o ller pointed out that some supplementary conditions on the
tetrads are needed. He suggested these conditions to arise from
suitable boundary conditions for the field equations,
possibly in the form of an anti-symmetric tensor.
These supplementary conditions would
uniquely determine a {\it tetrad lattice}\cite{Mol},
apart from a constant rotation of the tetrads in the lattice
The problem of consistently defining these supplementary
conditions is likely to be related to the symmetry
transformation determined by (53).

The Hamiltonian density (52) determines
the time evolution of field quantities via equation
(48), and in particular of the metric tensor $g_{ij}$
of three-dimensional spacelike hypersurfaces. This
property might simplify approaches to a canonical,
nonperturbative quantization of gravity provided we manage
to construct the reduced phase space determined by (36).

After implementing the primary constraints via
equations (36), the first term of $C^a$ is given by
$-\partial_i \Pi^{ai}$, with $\Pi^{ai}$ defined by (9).
From our previous experience (cf. ref.
\cite{Maluf2}) we are led to conclude that this term is
related to energy and momentum of the gravitational
field. In the present case we also interpret equations
$C^a=0$ as energy-momentum equations for the gravitational
field. According to this interpretation, 
the integral form of the constraint
equation $C^{(0)}=0$ can be written in the form
$E-{\cal H}=0$. Integration of $-\partial_i \Pi^{ai}$ over the
whole three-dimensional space yields the ADM energy.
A complete analysis of this issue will be presented elsewhere.

\bigskip
\noindent {\it Acknowledgements}\par
\noindent J. F. R. N. is supported by CAPES, Brazil.\par

\end{document}